\documentstyle[12pt,epsfig]{article}
\makeatletter
\def\@cite#1#2{{[{#1}]\if@tempswa\typeout
{IJCGA warning: optional citation argument
ignored: `#2'} \fi}}

\newcount\@tempcntc
\def\@citex[#1]#2{\if@filesw\immediate\write\@auxout{\string\citation{#2}}\fi
  \@tempcnta\z@\@tempcntb\m@ne\def\@citea{}\@cite{\@for\@citeb:=#2\do
    {\@ifundefined
       {b@\@citeb}{\@citeo\@tempcntb\m@ne\@citea\def\@citea{,}{\bf ?}\@warning
       {Citation `\@citeb' on page \thepage \space undefined}}%
    {\setbox\z@\hbox{\global\@tempcntc0\csname b@\@citeb\endcsname\relax}%
     \ifnum\@tempcntc=\z@ \@citeo\@tempcntb\m@ne
       \@citea\def\@citea{,}\hbox{\csname b@\@citeb\endcsname}%
     \else
      \advance\@tempcntb\@ne
      \ifnum\@tempcntb=\@tempcntc
      \else\advance\@tempcntb\m@ne\@citeo
      \@tempcnta\@tempcntc\@tempcntb\@tempcntc\fi\fi}}\@citeo}{#1}}
\def\@citeo{\ifnum\@tempcnta>\@tempcntb\else\@citea\def\@citea{,}%
  \ifnum\@tempcnta=\@tempcntb\the\@tempcnta\else
   {\advance\@tempcnta\@ne\ifnum\@tempcnta=\@tempcntb \else \def\@citea{--}\fi
    \advance\@tempcnta\m@ne\the\@tempcnta\@citea\the\@tempcntb}\fi\fi}
\makeatother
\newenvironment{Eqnarray}%
     {\arraycolsep 0.14em\begin{eqnarray}}{\end{eqnarray}}

\def\be{\begin{equation}}
\def\ee{\end{equation}}
\def\bear{\be\begin{array}}
\def\eear{\end{array}\ee}
\def\bea{\begin{Eqnarray}}
\def\eea{\end{Eqnarray}}

\def\lsim{\mathrel{\raise.3ex\hbox{$<$\kern-.75em\lower1ex\hbox{$\sim$}}}}
\def\gsim{\mathrel{\raise.3ex\hbox{$>$\kern-.75em\lower1ex\hbox{$\sim$}}}}
\def\ifmath#1{\relax\ifmmode #1\else $#1$\fi}
\def\ls#1{\ifmath{_{\lower1.5pt\hbox{$\scriptstyle #1$}}}}

\def\beq{\begin{equation}}
\def\eeq{\end{equation}}
\def\beqa{\begin{Eqnarray}}
\def\eeqa{\end{Eqnarray}}

\def\gappeq{\mathrel{\rlap {\raise.5ex\hbox{$>$}}
{\lower.5ex\hbox{$\sim$}}}}
\def\lappeq{\mathrel{\rlap{\raise.5ex\hbox{$<$}}
{\lower.5ex\hbox{$\sim$}}}}

\begin{document}
\def\IJMPA #1 #2 #3 {{\sl Int.~J.~Mod.~Phys.}~{\bf A#1}\ (19#2) #3$\,$}
\def\MPLA #1 #2 #3 {{\sl Mod.~Phys.~Lett.}~{\bf A#1}\ (19#2) #3$\,$}
\def\NPB #1 #2 #3 {{\sl Nucl.~Phys.}~{\bf B#1}\ (19#2) #3$\,$}
\def\PLB #1 #2 #3 {{\sl Phys.~Lett.}~{\bf B#1}\ (19#2) #3$\,$}
\def\PR #1 #2 #3 {{\sl Phys.~Rep.}~{\bf#1}\ (19#2) #3$\,$}
\def\JHEP #1 #2 #3 {{\sl JHEP}~{\bf #1}~(19#2)~#3$\,$}
\def\PRD #1 #2 #3 {{\sl Phys.~Rev.}~{\bf D#1}\ (19#2) #3$\,$}
\def\PTP #1 #2 #3 {{\sl Prog.~Theor.~Phys.}~{\bf #1}\ (19#2) #3$\,$}
\def\PRL #1 #2 #3 {{\sl Phys.~Rev.~Lett.}~{\bf#1}\ (19#2) #3$\,$}
\def\RMP #1 #2 #3 {{\sl Rev.~Mod.~Phys.}~{\bf#1}\ (19#2) #3$\,$}
\def\ZPC #1 #2 #3 {{\sl Z.~Phys.}~{\bf C#1}\ (19#2) #3$\,$}
\def\PPNP#1 #2 #3 {{\sl Prog. Part. Nucl. Phys. }{\bf #1} (#2) #3$\,$}

\catcode`@=11
\newtoks\@stequation
\def\subequations{\refstepcounter{equation}%
\edef\@savedequation{\the\c@equation}%
  \@stequation=\expandafter{\theequation}
  \edef\@savedtheequation{\the\@stequation}
  \edef\oldtheequation{\theequation}%
  \setcounter{equation}{0}%
  \def\theequation{\oldtheequation\alph{equation}}}
\def\endsubequations{\setcounter{equation}{\@savedequation}%
  \@stequation=\expandafter{\@savedtheequation}%
  \edef\theequation{\the\@stequation}\global\@ignoretrue

\noindent}
\catcode`@=12

\title{\hspace{4.1in}{\small CERN-TH/2003-294}\\\hspace{4.1in}{\small OUTP 0333P}\bigskip \\
Neutrino Phenomenology -- the case of two right handed neutrinos. }
\author{A. Ibarra$^{a}$ and G. G. Ross$^{a,b}$ \\
$^{a}$ Theory Division, CERN, CH-1211 Geneva 23, Switzerland\\
$^{b}$Department of Physics, Theoretical Physics, University of Oxford,\\
1 Keble Road, Oxford OX1 3NP, U.K.}
\date{}
\maketitle

\begin{abstract}
We make a general analysis of neutrino phenomenology for the case neutrino
masses are generated by the see-saw mechanism with just two right handed
neutrinos. We find general constraints on leptogenesis and lepton flavour
violating processes. We also analyse the predictions following from a
nontrivial texture zero structure.
\end{abstract}

\section{Introduction}
The see-saw mechanism \cite{seesaw} for generating neutrino 
masses remains the most plausible one to describe 
the observed neutrino masses. It relates the
smallness of the observed masses to the inverse of the mass scale at which
the strong weak and electromagnetic couplings unify. Such a structure is
natural if one extends the Standard Model to include Standard Model singlet
right handed neutrino states, something that recovers the quark lepton
symmetry.

However the analysis of the neutrino phenomenology coming from the see-saw
mechanism is made more difficult because it involves many more parameters
than can be measured from the neutrino masses and mixings. One may hope that
the observed structure will reveal a simplification and that the full range
of parameters is not necessary. For example it may be that an underlying
symmetry relates the fundamental Yukawa couplings or elements of the
Majorana mass matrix or forbids the appearance of one or more of them --
\textquotedblleft texture zeros\textquotedblright . Another possibility that
has been explored recently in 
\cite{Frampton:2002qc}
is that only two of the right-handed neutrinos (2RHN) play a role in the
determination of neutrino properties \cite{related}. This also reduces the number of free
parameters.

In order to look for the phenomenological implications of such structure it
is necessary to systemetize the parameterisation of the see saw mechanism.
In \cite{Ibarra:2003xp} it was shown how this could be done using a
parameterisation suggested by \cite{Casas:2001sr}. The method was applied to
the case that there were additional texture zeros limiting the number of
free parameters. In this paper we extend this analysis by exploring the
phenomenological implications of the 2RHN case. We start in Section \ref%
{count} with a discussion of the number of free parameters in this case and
in Section \ref{2rhn} we determine how the general parameterisation of \cite%
{Ibarra:2003xp} is modified. Section \ref{lept} shows that, despite the fact
that there are still two undetermined parameters, there are general
constraints on leptogenesis and lepton flavour violation in the 2RHN case. To
go further requires some model dependent reduction in the number of free
parameters. In Section \ref{tz} we discuss how this comes about if the
Yukawa couplings have one or more texture zero, presumably originating from
an underlying symmetry. Section \ref{1tz} determines the predictions that
result if there is one texture zero and Section \ref{2tz} does the same for
the two texture zero case. Finally Section \ref{conc} gives our conclusions.

\section{Parameter counting for the two right handed neutrino case.\label%
{count}}

If only two right handed neutrinos play a role in the see-saw mechanism
there is a reduction in the number of parameters needed. To see this we
consider first the case of three generations of left-handed $SU(2)$ doublet
neutrinos, $\nu _{L,i,}$ and three generations of right-handed Standard
Model singlet neutrinos, $\nu _{R,i}$ (3RHN). The Lagrangian responsible for
lepton masses has the form%
\begin{equation}
{\cal L}_{lep}={\nu_{R}^{c}}^T {\bf Y}_{\nu}~\nu _{L}\langle H^{0}\rangle
+{l_{R}^{c}}^T{\bf Y}_l~l_{L}\langle \overline{H}^{0}\rangle -
\frac{1}{2}{\nu_{R}^{c}}^T {\bf M_{\nu}}~\nu _{R}^{c},  \label{mass}
\end{equation}%
where ${\bf Y}_{\nu}$ and ${\bf Y}_{l} $ are the 
matrices of Yukawa couplings which
give rise to the neutrino and charged lepton Dirac mass matrices
respectively and ${\bf M_{\nu}}$ is the neutrino Majorana mass matrix. The
light neutrino mass matrix, ${\cal M}{\it ,}$ is given by the see-saw form%
\begin{equation}
{\cal M}={\bf Y}_{\nu}^{T}~{\bf M^{-1}_{\nu}}~{\bf Y}_{\nu}, \label{seesaw}
\end{equation}%
Consider the basis in which the Majorana masses and the charged leptons are
diagonal and real. In this case there are 3 Majorana masses together with 18
real parameters (9 angles and 9 phases) needed to specify ${\bf Y}_{\nu }$. 
Of these, 3 phases are unphysical and can be eliminated 
by a redefinition of the
left handed lepton doublet. However not all of the remaining parameters are
independent in the way they determine ${\cal M}.$ In particular, in the
basis in which ${\bf M_{\nu}}$ is diagonal, $D_{M_{\nu}}=diag(M_1,M_2,M_3)$,
 a simultaneous rescaling of $%
\left(\bf{Y_{\nu }}\right) _{Ij},$ $j=1,2,3$ with $I$ fixed by $\lambda $
can be absorbed by a rescaling of 
$M_{I}$ by $%
\lambda ^{2}$ so that the Majorana masses do not introduce additional
parameters. As a result there are only 15 effective parameters determining
the light neutrino mass matrix via the see-saw mechanism. The measureable
parameters associated with the light neutrino mass matrix consist of 3
masses plus three mixing angles and 3 phases, a total of 9 measureables.
That means 6 parameters associated with the see-saw mechanism are not
determined by the neutrino masses, mixing angles and phases. In \cite%
{Casas:2001sr} a general parameterisation of these parameters was given. In
it the most general neutrino Yukawa coupling which is compatible with low
energy data, written in the basis where the charged lepton Yukawa coupling
and the right-handed mass matrix are diagonal, is given by 
\begin{equation}
{\bf Y}_{\nu}=D_{\sqrt{{\bf M_{\nu}}}}RD_{\sqrt{m}}U^{\dagger }/\langle
H^0\rangle,   \label{ibca}
\end{equation}%
where $D_{\sqrt{{\bf M_{\nu}}}}$ is the diagonal
matrix of the square roots of the
eigenvalues of ${\bf M_{\nu}}$, $D_{\sqrt{m}}$ is the diagonal matrix of the
roots of the physical masses, $m_{i},$ of the light neutrinos, $U$ is the
Maki-Nakagawa-Sakata (MNS) matrix \cite{Maki:mu} and $R$ is a $3\times 3$ 
orthogonal matrix which parameterises the
information that is lost in the decoupling of all three right-handed
neutrinos. Notice that we have included all the low
energy phases in the definition of the matrix $U$, 
{\it i.e.} we have written the MNS matrix in the 
form $U=V~{\rm diag}~(e^{-i\phi /2},e^{-i\phi ^{\prime }/2},1) $,
where $\phi $ and $\phi ^{\prime }$ are the CP violating phases and $V$ has
the form of the CKM matrix. It is important to note
that $R$ can be complex as long as $R^{T}R=RR^{T}=1$. Thus $R$ has 6 real
parameters corresponding to the 6 undetermined parameters discussed above.
The mass matrix, ${\cal M}$, is determined by 
${\bf Y}_{\nu}D^{-1}_{\sqrt{{\bf M_{\nu}}}}$ and, 
as expected, it is not separately dependent on ${\bf M_{\nu}}$.

If only 2RHN contribute to the see-saw mechanism the number of parameters is
reduced. In this case only 12 real parameters (6 moduli and 6 phases) are
needed to specify $\left({\bf Y}_{\nu}\right) _{ij},$ $i=1,2,$ $j=1,2,3.$
Allowing for 3 redundant phases and the rescaling of the 2 Majorana masses
the effective number is reduced to 7 plus, of course, the 2 Majorana masses.
The number of measureable parameters is only reduced by the 2 corresponding
to one mass and one phase. The conclusion is that in the 2RHN case there are
only (9-7)=2 real parameters determining the light neutrino mass matrix via
the see-saw mechanism.

\section{The 2RHN model as the decoupling limit of the 3RHN model\label{2rhn}}

The 2RHN model can be regarded as the limiting case of the three
right-handed neutrino model in which one of the right-handed neutrinos has
an infinite mass, while all the Yukawa couplings remain perturbative. As we
have discussed, the 2RHN model depends on 4 fewer parameters than the 3RHN
model, and this should be reflected in the number of parameters of the
matrix $R$. This may be seen by taking the limit in which one of the right
handed neutrinos has an infinite mass, say $M_{3}.$ In this case two angles
in the matrix $R$ are determined. The reason is the following. From eq.(\ref%
{ibca}) one finds that the elements of the third row of $R$ are given by 
\begin{equation}
R_{3i}=\frac{({\bf Y}_{\nu}U)_{3i}}{\sqrt{M_{3} m_{i}}}\langle H^0\rangle.
\end{equation}%
Since the numerator is finite and $m_{2}$ and $m_{3}$ are different from
zero, $R_{32}$ and $R_{33}$ have to vanish as $M_{3}$ goes to infinity. On
the other hand, $m_{1}\rightarrow 0$ as $M_{3}\rightarrow \infty $, so the
limit of $R_{31}$ is not well defined and might be non-zero. However, the
orthogonality of $R$ requires $R_{31}$ to be unity and $R_{i1}=0$, $i=1,2$.
Therefore, in the limit $M_{3}\rightarrow \infty $, the matrix $R$ takes the
form: 
\begin{equation}
R=\left( 
\begin{array}{ccc}
0 & \cos z & \pm \sin z \\ 
0 & -\sin z & \pm \cos z \\ 
1 & 0 & 0%
\end{array}%
\right),  \label{R3x3}
\end{equation}%
where $z$ is a complex angle and the $\pm $ in the third column has been
included to account for the possible reflections in the orthogonal matrix $R$.
It is clear that in the 2RHN model, the corresponding matrix $R$ is simply
given by the first two rows of eq.(\ref{R3x3}).

Using this form for $R$ the different elements of the neutrino Yukawa matrix
read: 
\begin{eqnarray}
{{\bf Y}_{\nu}}_{1i} &=&\sqrt{M_{1}}(\sqrt{m_{2}}\cos z~U_{i2}^{\ast}\pm 
\sqrt{m_{3}}\sin z~U_{i3}^{\ast })/\langle H^0 \rangle,  \label{y} \\
{{\bf Y}_{\nu}}_{2i} &=&\sqrt{M_{2}}(-\sqrt{m_{2}}\sin z~U_{i2}^{\ast }\pm 
\sqrt{m_{3}}\cos z~U_{i3}^{\ast })/\langle H^0 \rangle,  \nonumber 
\end{eqnarray}%
where $i=1,2,3$. The unknown complex parameter $z$ encodes the real
parameter and the phase necessary to match the total number of parameters at
high energies and at low energies in the 2RHN model.

A word of caution is in order concerning the relation between the decoupling
limit of the 3RHN case and the 2RHN case. In the 3RHN case the Yukawa
couplings ${{\bf Y}_{\nu}}_{3i}$ are not necessarily negligible due to the
factor $\sqrt{M_{3}}$ appearing in eq.(\ref{ibca}), and could produce some
effect at low energies through the radiative corrections. 
Whether or not they are,
depends on the magnitude of the elements $R_{3j},$ $j=2,3$ which vanish
like $1/\sqrt{M_{3}}.$ Moreover, the decoupling limit 
corresponds to the case that the
third Majorana mass is at, or above, the cutoff scale so that the third
neutrino does not contribute, even via radiative corrections.

\section{Leptogenesis and lepton flavour violation in the 2RHN case.\label%
{lept}}

Without making assumptions about the parameters relevant at high energy
scales there are no definite predictions for the low energy neutrino
parameters. However, the reduction in the number of unknown parameters with
respect to the 3RHN model makes it possible to extract some general features
of this case, in particular, for thermal leptogenesis \cite{Fukugita:1986hr}
and for rare lepton decays induced by radiative corrections 
\cite{Borzumati:1986qx}.

\subsection{Thermal leptogenesis}

First, we derive some general constraints on the thermal leptogenesis scenario,
assuming that the decay of the lightest right-handed neutrino is the only
source of lepton asymmetry \cite{Buchmuller:2003gz}. 
If this is the case, the CP asymmetry produced
by heavy lepton decay can be written as (we will drop small
correction terms of $O(m_2/m_3)$)
\begin{equation}
\frac{\epsilon }{|\epsilon _{\max }|}\simeq -\frac{Im(\sin ^{2}z)}{
\left\vert \sin ^{2}z\right\vert +\frac{m_{2}}{m_{3}}\left\vert \cos
^{2}z\right\vert },  \label{epsilon}
\end{equation}
where $|\epsilon _{\max }|\simeq \frac{3}{8\pi }\frac{M_{1}m_{3}}
{\langle H^0\rangle ^{2}}$ \cite{Davidson:2002qv}. It is clear that for a
suitable choice of $z$ it is possible to maximise $\epsilon .$ However
thermal leptogenesis can only lead to an acceptable level of baryogenesis if
the subsequent washout effects are not too large. They can be conveniently
characterized by \cite{Plumacher:1996kc} the parameter $\widetilde{m}_{1}$,
the parameter being bounded above if washout processes are to be limited. In
the 2RHN case. 
\begin{equation}
\widetilde{m}_{1}=m_{2}\left\vert \cos ^{2}z\right\vert +m_{3}\left\vert
\sin ^{2}z\right\vert.   \label{mtilde}
\end{equation}%
Notice that $\widetilde{m}_{1}\geq m_2$ and so there 
is a potential conflict with the upper bound following from the
condition that washout is acceptable \cite{Ibarra:2003xp},
\cite{Chankowski:2003rr}. To quantify this we note that from
eqs.(\ref{epsilon}) and (\ref{mtilde}) one finds 
\begin{equation}
\left\vert \frac{\epsilon}{\epsilon_{max}}\right\vert\lsim 
(1-\frac{m_2}{\widetilde{m}_{1}}).   \label{eps-mtilde}
\end{equation}%
The problem is now obvious because, if the washout effects are minimal 
(minimum $\widetilde{m}_{1}$ implying from eq.(\ref{mtilde}) that 
$\widetilde{m}_{1}=m_{2}$), the CP asymmetry vanishes. Physically, taking $%
\widetilde{m}_{1}\simeq m_{2}$ corresponds to the case in which the heaviest
light neutrino state is dominated by the lightest right-handed neutrino
state ($\cos z\simeq 1$). On the other hand, when the CP asymmetry is close
to maximal, $\widetilde{m}_{1}$ must be very large. In this case, 
to avoid washout \cite{Plumacher:1996kc}, we must have a large 
right handed neutrino mass, $M_{1}\ge 10^{11}$ GeV. 
In particular, this is the case for the limit in which the
heaviest light neutrino state is dominated by the heaviest right-handed
neutrino state ($\cos z\simeq 0$). In a supersymmetric model with gravity
mediated supersymmetry breaking, the reheat temperature 
which is related to $M_1$ is high and could be in conflict with the
gravitino overproduction constraints. However,
this problem can be circumvented in other supersymmetry breaking mediation
scenarios, such as gauge mediation, where the gravitino can be much lighter 
or through a weakening of the constraints on the reheat temperature 
\cite{Buchmuller}.

\subsection{Rare processes}

The general 2RHN structure also has implications for rare processes 
in supersymmetric scenarios. We concentrate in the most conservative 
case from the point of flavour violation\footnote{Therefore, our results 
should be understood as lower bounds on the  rates for these rare 
processes, barring cancellations among different contributions.}, 
namely the class of scenarios where supersymmetry is broken in
a hidden sector, and the breaking is transmitted to the observable sector by
a flavour blind mechanism, like gravity. If this is the case, all
the soft breaking terms are diagonal at the high energy scale, and the only
source of flavour violation in the leptonic sector 
are the radiative corrections to the soft terms, through the neutrino Yukawa 
couplings. In this class of scenarios,
the rate for the process $l_i\rightarrow l_j \gamma$ is given by 
\cite{Casas:2001sr}
\begin{equation}
BR(l_i\rightarrow l_j \gamma) \simeq \frac{\alpha^3}{G_F^2 m_S^8}
[\frac{1}{8 \pi^2}(3 m^2_0 +A^2_0)]^2 |C_{ij}|^2 \tan^2 \beta,
\label{BR}
\end{equation}
where $C_{ij}=({\bf Y}_{\nu}^{\dagger }\log \frac{M_{X}}{M}
{\bf Y}_{\nu})_{ij}$ is the crucial
quantity to determine the size of these rates. In this formula, 
$m_S$ represents supersymmetric leptonic masses, and $m_0$ and $A_0$
are, respectively, the universal scalar soft mass 
and the trilinear term at the GUT scale.
If all
 the parameters are real, we estimate these quantities to be $%
|C_{12}|\le {\cal O}(0.1)\frac{\sqrt{m_{2}m_{3}}}{\langle H\rangle ^{2}}%
M_{2}\log \frac{M_{X}}{M_{2}}$ and $|C_{13,23}|\sim \sqrt{\frac{m_{3}}{m_{2}}%
}C_{12}$. The predicted rates for the rare processes are therefore rather
small, unless $M_{2}$ is large. For instance, from the present bound on the
process $\mu \rightarrow e\gamma $ one can set the bound $M_{2}\leq 10^{14}$
GeV. This bound could be improved by one order of magnitude if the next
generation of experiments reach the projected sensitivity, BR$(\mu
\rightarrow e\gamma )\le 10^{-13}$.

\begin{figure}[tbp]
\centerline{
\psfig{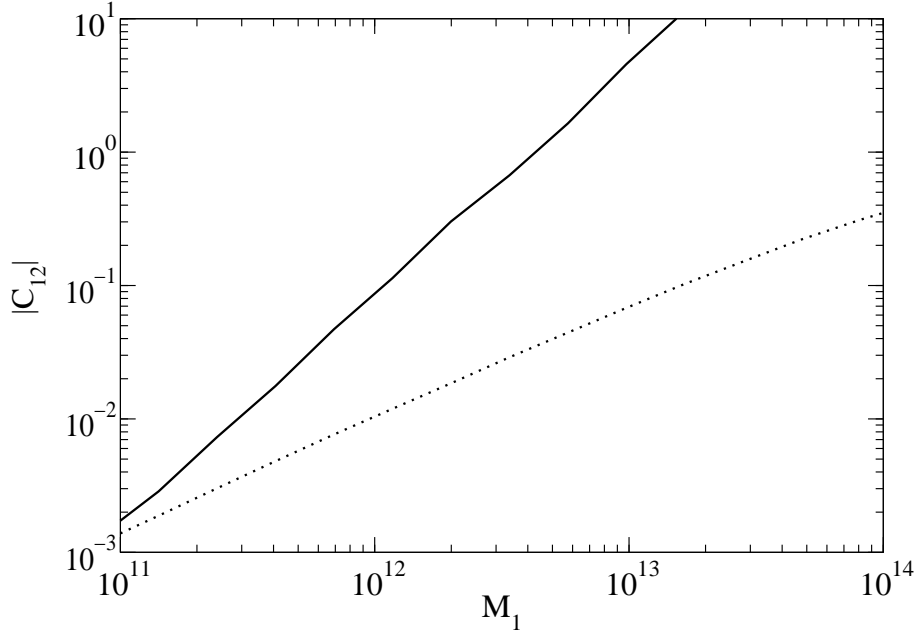}}
\caption{\protect\footnotesize Maximum value of the parameter 
$|C_{12}|=|({\bf Y}_{\nu}^{\dagger }\log \frac{M_{X}}{M}{\bf Y}_{\nu})_{12}|$
for all the see-saw scenarios that reproduce the observed masses and mixing
angles, when there is no CP violation (dotted line) and when the CP
violation is consistent with the baryon asymmetry of the Universe through
the mechanism of leptogenesis. In this plot we have set $U_{13}=0$ and
$M_2=10 M_1$.}
\label{fig1}
\end{figure}

However, if leptogenesis is the correct mechanism to generate the baryon
asymmetry of the Universe, the parameter $z$ has to be complex and this
could enhance the rates for the rare processes. We concentrate on the
combination $C_{12}$ that is related to the process $\mu \rightarrow e\gamma$. 
In Figure \ref{fig1} we show the maximum value of $|C_{12}|$  
as a function of the lightest right-handed
neutrino mass, $M_1$, assuming $U_{13}=0$ and $M_2=10 M_1$. For
other hierarchies of right-handed masses, 
the results scale roughly as $(M_2/M_1)^2$. In this plot, 
we take random values for  $\phi'$ and $z$, and we fix the 
lightest right-handed neutrino mass by requiring a 
correct baryon asymmetry $\eta_B\simeq 6\times 10^{-10}$,
as reported by the WMAP collaboration. To compute the baryon asymmetry,
we used the approximate treatment of the dilution effects described
in \cite{Nielsen:2001fy}. In the plot we have also used 
the best fit points for the solar angle and the neutrino masses
reported in \cite{Maltoni:2003da}, and we have 
assumed a hierarchical spectrum of neutrino
masses. The enhancement in the rates after including the constrains from
leptogenesis illustrates the interesting interplay between low energy lepton
flavour violation and leptogenesis in supersymmetric scenarios.

\section{Parametrization of the see-saw mechanism in the texture zero basis 
\label{tz}}

As was discussed in \cite{Ibarra:2003xp}, further assumptions about the
physics relevant at a high energy scale can reduce the number of parameters
and even lead to relations among the low energy observables. In the
remainder of this paper we consider the possibility that one or more
elements of the Yukawa coupling matrix are anomalously small and can be
ignored, the so-called \textquotedblleft texture zeros\textquotedblright .
For example, one texture zero in the neutrino Yukawa matrix would fix the
matrix $R$ up to \textquotedblleft reflections\textquotedblright , and two
texture zeros would lead to relations among the mixing angles and the
neutrino masses. One reason for the interest in texture zeros is that they
may indicate the presence of a new family symmetry which require certain
matrix elements be anomalously small. Thus identification of texture zeros
may be an important step in unravelling the origin of the fermion masses and
mixings. It is already known that the measured quark masses and mixing
angles are consistent with such texture zeros \cite{Ramond:1993kv},\cite%
{Roberts:2001zy}. In this and the following sections we will explore the
different consequences at low energies of the 2RHN model assuming that there
are texture zeros.

In general, texture zeros do not appear in the basis where the charged
lepton Yukawa couplings and the right-handed mass matrix are simultaneously
diagonal. To allow for this we write the Lagrangian responsible for 
lepton masses in the texture zero basis as: 
\begin{equation}
{\cal L}_{lep}^{TZ}={\nu _{R}^{c}}^T {\bf Y}_{\nu}^{\bf TZ}
~\nu _{L}\langle H^{0}\rangle
+{l_{R}^{c}}^T{\bf Y}_l^{\bf TZ} ~l_{L}\langle \overline{H}^{0}\rangle -
\frac{1}{2}{\nu_{R}^{c}}^T {\bf M_{\nu}^{TZ}}~\nu _{R}^{c}.  
\end{equation}%
We can diagonalize the charged lepton Yukawa matrix by ${\bf Y}_l^{\bf TZ}%
=V_{l}D_{{\bf Y}_l}U_{l}^{\dagger }$ and the right-handed mass matrix by 
${\bf M_{\nu}^{TZ}}=V_{\nu }^{\ast }D_{\bf M_{\nu}}V_{\nu }^{\dagger }$. 
Then, the most general neutrino Yukawa coupling that is 
compatible with the low-energy data, written in the texture zero basis, 
is given by: 
\begin{equation}
{\bf Y}_{\nu}^{\bf TZ}=V_{\nu }^{\ast }D_{\sqrt{{\bf M_{\nu}}}}
R D_{\sqrt{m}}W^{\dagger }/\langle H^{0}\rangle,  \label{ibro}
\end{equation}
where $W=U_{l}U$. Here, $R$ is 
\begin{equation}
R=\left( 
\begin{array}{ccc}
0 & \cos z & \pm \sin z \\ 
0 & -\sin z & \pm \cos z%
\end{array}%
\right). 
\end{equation}

In the texture zero basis the right-handed mass matrix can be diagonalized
by a unitary matrix that in general depends on three phases and one
rotation angle. One of these phases cannot be removed by redefinitions of
the right-handed neutrino fields, due to the Majorana nature of these
particles. On the other hand, the other two can be removed without altering
the texture zero structure of the Yukawa matrices. Hence, in our texture
zero basis, the unitary matrix $V_{\nu }$ can be parametrized by 
\begin{equation}
V_{\nu }=\left( 
\begin{array}{cc}
{\bf \cos \omega } & {\bf \sin \omega } \\ 
-{\bf \sin \omega } & {\bf \cos \omega }%
\end{array}%
\right) .\left( 
\begin{array}{cc}
1 &  \\ 
& e^{-i\alpha }%
\end{array}%
\right), 
\end{equation}
where $\omega $ is the mixing angle, and $\alpha $ is the Majorana phase.

There remains the question of the form of the lepton mixing matrix in the
texture basis, needed to determine $W.$ The form of this has been discussed
in \cite{Ibarra:2003xp}. For the case the lepton mass matrix has off diagonal
elements whose magnitude is approximately symmetric and that, like the
quarks, the hierarchy of lepton masses is due to an hierarchical structure
in the matrix elements and not due to a cancellation between different
contributions one has the bounds 
\begin{eqnarray}
\left\vert \left( U_{l}\right) _{23}\right\vert &\leq &\sqrt{\frac{m_{\mu }}{%
m_{\tau }}}, \nonumber\\
\left\vert \left( U_{l}\right) _{12}\right\vert &\leq &\sqrt{\frac{m_{e}}{%
m_{\mu }}}, \nonumber\\
\left\vert \left( U_{l}\right) _{13}\right\vert &\leq &\sqrt{\frac{m_{e}}{%
m_{\tau }}}.
\label{lepton}
\end{eqnarray}%
In addition, it is necessary to determine the phases 
in $U_{l}$. There is a residual phase ambiguity because the basis in
which the MNS matrix has the standard form can be different from the
''symmetry" basis in which the texture zero appears. This corresponds to the
simultaneous redefinition of the phase of the left- and right- handed
states such that the Dirac structure is invariant. 
With this we have $W$ $=|U_{l}|PU$ where 
$P=diag(e^{i\alpha _{1}},e^{i\alpha _{2}},e^{i\alpha _{3}})$.
In practice the magnitudes of $\left( U_{l}\right) _{23}$ and $\left(
U_{l}\right) _{13}$ are so small that they do not affect the mixing coming
from the neutrino sector. However $\left( U_{l}\right) _{12}$ close to the
upper bound given in eq.(\ref{lepton}) does give a significant contribution to
the CHOOZ angle. Its effect is considered below. If the magnitude of the
charged lepton mass matrix elements departs significantly from the symmetric
form there is no constraint on the magnitude of the matrix elements of $%
U_{l}.$ In this case the contributions to the MNS matrix coming from the
neutrino sector should be considered as an indication of the lower bound on
the MNS matrix elements, assuming there is no delicate cancellation
between the contributions of $U_{l}$ and $U_{\nu }.$

With this parametrization of the see-saw mechanism, it is straightforward to
compute predictions at low energies from texture zeros. One texture zero in
the texture zero basis would fix the angle, $z$, in $R$; two texture zeros
would yield relations among physical parameters (light neutrino masses,
mixing angles in $W$ and right handed sector parameters). Although some of
these physical parameters are not measurable, for instance the right handed
sector parameters, this approach will allow us to predict ranges for the
measurable quantities, in particular the CHOOZ angle.

\section{Predictions from models with one texture zero\label{1tz}}

\begin{figure}[tbp]
\centerline{
\psfig{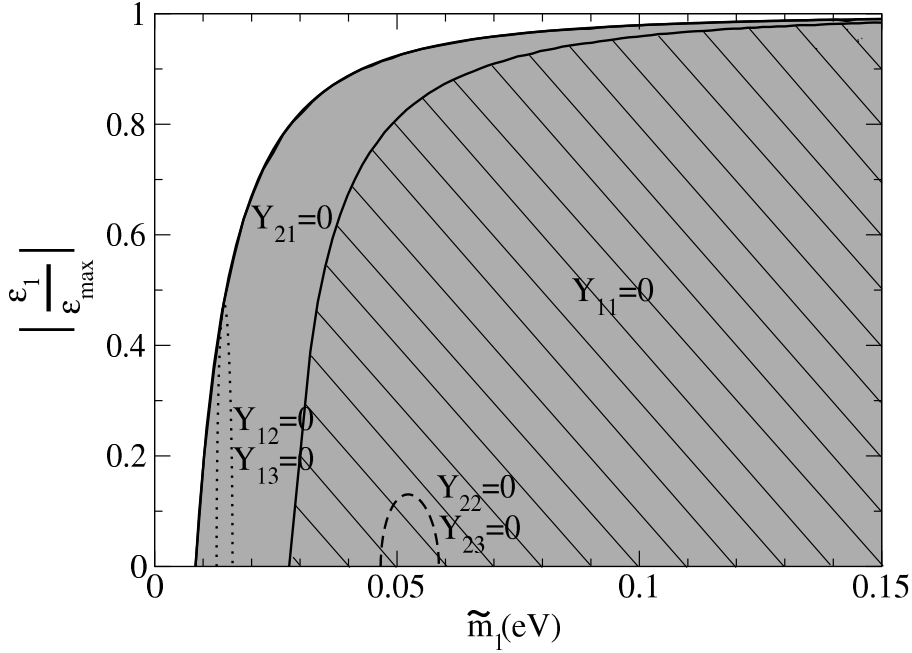}}
\caption{\protect\footnotesize Allowed regions in the plane 
$\widetilde{m}_{1}-|\frac{\epsilon _{1}}{\epsilon _{max}}|$,
for the neutrino Yukawa couplings with one texture zero 
that are compatible with all the available data at low energies.
The shaded area applies to the case $Y_{21}=0$ and the
hatched area to the case $Y_{11}=0$. The dotted line shows 
the allowed area for the cases $Y_{12}=0$ or $Y_{13}=0$ and 
the dashed line shows the allowed area for the cases $Y_{22}=0$ or $Y_{23}=0$.
In this plot we have assumed that the texture zero appears in the basis
where the charged lepton Yukawa coupling and the 
right-handed mass matrix are both diagonal.}
\label{fig2}
\end{figure}

One texture zero in the neutrino Yukawa coupling fixes the unknown parameter 
$z$. From eq.(\ref{y}), we obtain 
\begin{equation}
\tan z=\mp \sqrt{\frac{m_{2}}{m_{3}}}.\frac{W_{i2}^{\ast }\pm \sqrt{\frac{%
M_{2}}{M_{1}}}\sqrt{\frac{m_{3}}{m_{2}}}W_{i3}^{\ast }e^{i\alpha }\tan
\omega }{W_{i3}^{\ast }\mp \sqrt{\frac{M_{2}}{M_{1}}}\sqrt{\frac{m_{2}}{m_{3}%
}}W_{i2}^{\ast }e^{i\alpha }\tan \omega },  \label{tz1}
\end{equation}%
when ${\bf Y}_{1i}^{\bf TZ}=0$, and 
\begin{equation}
\tan z=\pm \sqrt{\frac{m_{3}}{m_{2}}}~\frac{W_{i3}^{\ast }\mp \sqrt{\frac{%
M_{1}}{M_{2}}}\sqrt{\frac{m_{2}}{m_{3}}}W_{i2}^{\ast }e^{-i\alpha }\tan
\omega }{W_{i2}^{\ast }\pm \sqrt{\frac{M_{1}}{M_{2}}}\sqrt{\frac{m_{3}}{m_{2}%
}}W_{i3}^{\ast }e^{-i\alpha }\tan \omega },  \label{tz2}
\end{equation}%
when ${\bf Y}_{2i}^{\bf TZ}=0$. With only these hypotheses, there are no predictions
for the low energy parameters. However, fixing $z$ imposes further
restrictions on the leptogenesis parameters $\epsilon $ and $\widetilde{m}%
_{1}$. These are summarized in Figure \ref{fig2}, where we show the allowed
regions in the plane $\widetilde{m}_{1}-|\frac{\epsilon _{1}}{\epsilon _{max}%
}|$. For the (1,1) and (2,1) texture zeros, $\tan z$ depends on the CHOOZ
angle, which has not been measured. So, in the plot we have taken the CHOOZ
angle between zero and 0.23, which is the $3\sigma $ bound from the global
analysis \cite{Maltoni:2003da}.

From the figure, we see that Yukawa couplings with zeros in the (2,2) and
(2,3) positions are very disfavoured from the point of view of leptogenesis.
They yield small CP asymmetries, $|\frac{\epsilon _{1}}{\epsilon _{max}}%
|\lsim \frac{m_{2}}{m_{3}}$, and large washout effects, $\widetilde{m}%
_{1}\simeq m_{3}$. Similarly for the Yukawa coupling with a texture zero in
the (1,1) position. On the other hand, for Yukawa couplings with a texture
zero in the (2,1) position the bound is almost identical to the model
independent bound $|\frac{\epsilon }{\epsilon _{max}}|\lsim 1-\frac{m_{2}%
}{\widetilde{m}_{1}}$. However, a stronger bound on the CHOOZ angle
will make the allowed region smaller at high values of $\widetilde{m}_{1}$.
Yukawa matrices with a (1,2) or (1,3) texture zero are also favoured from
the point of view of leptogenesis. They yield almost minimal washout
effects, $\widetilde{m}_{1}\simeq m_{2}(1+\cos ^{2}\theta _{12})$ and
relatively large CP asymmetries, $|\frac{\epsilon _{1}}{\epsilon _{max}}%
|\lsim \frac{\cos ^{2}\theta _{12}}{1+\cos ^{2}\theta _{12}}$.

In Figure \ref{fig2} we have neglected the contributions to the mixing from
the charged lepton sector and the right-handed Majorana sector. These
mixings do not qualitatively change the results. For example, for the case
for the texture zero in the (2,2) position we still find most of the points
concentrated in the region around the dashed line in Figure \ref{fig2}.
There are only a few points saturating the model independent bound, eq.(\ref%
{eps-mtilde}), that correspond to special choices of the right-handed
parameters.

An interesting issue that has been discussed extensively in the literature
concerns the connection of leptogenesis and low energy observables 
\cite{Davidson:2002em}. In particular
the correlation between the sign of the baryon asymmetry and the CP
violation at low energies has been discussed 
\cite{Frampton:2002qc}.
This connection is clear only when $U_{l}\simeq 1$ and $V_{\nu }\simeq 1$,
otherwise some unmeasurable parameters in the charged-lepton sector or the
right-handed sector enter into play. We find that one texture zero is enough to
establish such connection, since the sign of the CP asymmetry is determined
by minus the argument of $\tan ^{2}z$, which in turn is fixed with one
single texture zero. For instance, when the texture zero appears in the (1,2)
or (1,3) position, 
\begin{eqnarray}
\frac{\epsilon _{1}}{|\epsilon _{max}|} &\simeq &-\frac{\sin \phi ^{\prime }%
}{1+|\frac{W_{i3}}{W_{i2}}|^{2}}, \nonumber \\
\widetilde{m}_{1} &\simeq &m_{2}(1+|\frac{W_{i2}}{W_{i3}}|^{2}),
\end{eqnarray}%
where $i=2,3.$ On the other hand, when it appears in the (2,2) or (2,3)
position, 
\begin{eqnarray*}
\frac{\epsilon _{1}}{|\epsilon _{max}|} &\simeq &+\sin \phi ^{\prime }.%
\frac{m_{2}}{m_{3}}|\frac{W_{i2}}{W_{i3}}|^{2}, \\
\widetilde{m}_{1} &\simeq &m_{2}.
\end{eqnarray*}%
Finally, the case when the texture zero appears in the first column deserves
some more careful analysis, since it involves the CHOOZ angle which has not
been measured. The expressions are rather complicated, but can be readily
computed from eqs.(\ref{epsilon}, \ref{mtilde}) and eqs.(\ref{tz1}, \ref{tz2}%
). We just show the numerical results in Figure 3, where we plot
the CP asymmetry divided by $\sin (\phi ^{\prime }-2\delta )$, to show
better the connection between the sign of the CP asymmetry and the low
energy CP violation. The lepton asymmetry and $\sin (\phi ^{\prime }-2\delta
)$ have the same sign for the (2,1) texture zero and opposite for the (1,1)
texture zero (the sign of the baryon asymmetry is opposite to the sign of
the lepton asymmetry).

\begin{figure}[tbp]
\centerline{
\psfig{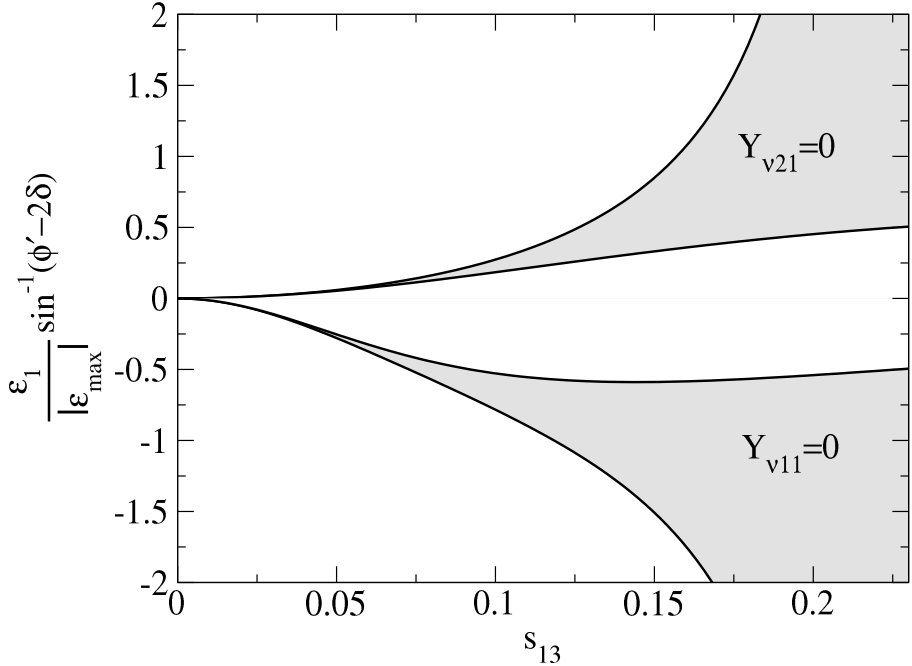}}
\label{Figure3}
\caption{\protect\footnotesize CP asymmetry divided
by the sine of a combination of phases relevant for the 
low energy CP violation, for different values of the CHOOZ angle. This
plot shows the correlation between the sign of the lepton asymmetry
and the sign of the low energy CP violation, when there is a texture
zero in the (1,1) or (2,1) position in the neutrino Yukawa coupling.}
\end{figure}


\section{Predictions from models with two texture zeros}\label{2tz}

We can write now the predictions from two texture zeros in the neutrino
Yukawa matrix. When the two texture zeros appear in the same row, 
${\bf Y}_{1i}^{\bf TZ}=0$, ${\bf Y}_{1j}^{\bf TZ}=0$  then 
\begin{equation}
\epsilon _{ijk}W_{k1}(e^{-i\alpha}M_1\cos^2\omega+
e^{i\alpha}M_2\sin^2\omega)=0.  \label{same-row1}
\end{equation}%
or if  ${\bf Y}_{2i}^{\bf TZ}=0$, ${\bf Y}_{2j}^{\bf TZ}=0$ then
\begin{equation}
\epsilon _{ijk}W_{k1}(e^{-i\alpha}M_1\sin^2\omega+
e^{i\alpha}M_2\cos^2\omega)=0.  \label{same-row2}
\end{equation}%

On the other hand, when the texture zeros appear in different rows, 
${\bf Y}_{1i}^{\bf TZ}=0$, ${\bf Y}_{2j}^{\bf TZ}=0$, 
the following relation holds: 
\begin{equation}
W_{i3}W_{j3}+\frac{m_{2}}{m_{3}}W_{i2}W_{j2}\pm \sqrt{\frac{m_{2}}{m_{3}}}%
\sin \omega \cos \omega ~(e^{i\alpha }\sqrt{\frac{M_{1}}{M_{2}}}-e^{-i\alpha
}\sqrt{\frac{M_{2}}{M_{1}}})\epsilon _{ijk}W_{k1}^{\ast }\det W=0,
\end{equation}%
where the $\pm $ comes from the two possible reflections in $R$, see eq.(\ref
{R3x3}).

Let us first discuss the results for the case where $U_{l}\simeq 1$ and $%
V_{\nu }\simeq 1$, i.e. when all the mixing in the neutrino sector comes
from the neutrino Yukawa matrix, and later on the general case, when there
are also contributions coming from the right-handed Majorana sector and the
charged-lepton sector.

\subsection{The case with $U_l \simeq 1$ and $V_{\protect\nu} \simeq 1$}

\begin{table}
\begin{center}
\begin{tabular}{ccccc} \hline
&Texture for ${\bf Y}_{\nu}$ && Predictions
\\ \hline \hline
I&$ \pmatrix{
 0     &    0   & \times \cr
\times & \times & \times}$, 
$\pmatrix{
\times & \times & \times \cr
 0     &    0   & \times}$ 
&& 
$U_{31}=0$
\\ \hline 
II&$ \pmatrix{
 0     & \times &   0   \cr
\times & \times & \times}$, 
$\pmatrix{
\times & \times & \times \cr
 0     & \times &   0}$ 
&& 
$U_{21}=0$
\\ \hline 
III&$ \pmatrix{
\times &    0   &   0   \cr
\times & \times & \times}$, 
$\pmatrix{
\times & \times & \times \cr
\times &    0   &   0   }$
&& 
$U_{11}=0$
\\ \hline 
IV&$ \pmatrix{
 0   & \times  & \times \cr
 0   & \times  & \times}$
&& 
$U_{13}\simeq \pm i \sqrt{\frac{m_2}{m_3}} \sin \theta_{sol}e^{-i \phi'/2}$
\\ \hline 
V&$ \pmatrix{
\times &    0   & \times \cr
\times &    0   & \times}$
&& 
$U_{23}\simeq \pm \sqrt{\frac{m_2}{2m_3}} \cos \theta_{sol}e^{-i \phi'/2}$
\\ \hline 

VI&$ \pmatrix{
\times & \times &  0  \cr
\times & \times &  0 }$
&& 
$U_{33}\simeq  \pm \sqrt{\frac{m_2}{2m_3}} \cos \theta_{sol}e^{-i \phi'/2}$
\\ \hline 
VII&$ \pmatrix{
 0     & \times  & \times \cr
\times &    0    & \times}$, 
$\pmatrix{
\times &    0    & \times \cr
 0     & \times  & \times}$ 
&& 
$U_{13}\simeq -\frac{m_2}{2m_3} \sin 2 \theta_{sol} e^{-i \phi'}$
\\ \hline 
VIII&$ \pmatrix{
 0     & \times  & \times \cr
\times & \times  &    0    }$, 
$\pmatrix{
\times & \times  &    0   \cr
 0     & \times  & \times}$ 
&& 
$U_{13}\simeq \frac{m_2}{2m_3} \sin 2 \theta_{sol} e^{-i \phi'}$
\\ \hline 
IX&$ \pmatrix{
\times &    0    & \times \cr
\times & \times  &    0    }$, 
$\pmatrix{
\times & \times  &    0   \cr
\times &    0    & \times}$ 
&& 
$U_{23}\simeq \frac{1}{\sqrt{2}}\frac{m_2}{m_3}
\cos^2 \theta_{sol} e^{-i \phi'}$
\\ \hline 
\end{tabular}
\end{center}
\caption{\protect\footnotesize Predictions following from the 
various two texture zero structures.}
\label{Table1}
\end{table}


Under these assumptions it is straightforward to compute the predictions at
low energies for all the fifteen possible Yukawa matrix with two texture
zeros\footnote{Some related analyses for this case can also be found in 
\cite{Frampton:2002qc}.}. 
These are summarized in Table \ref{Table1}, where we have set the
atmospheric angle to the experimentally favoured maximal value. Only five of
them are allowed by present experiments, namely textures IV, VII, and VIII.
The matrix with texture zeros in the
same column leads to the prediction for the CHOOZ angle $s_{13}\simeq \sqrt{%
\frac{m_{2}}{m_{3}}}\sin \theta _{sol}\simeq 0.22$, which is marginally
allowed, and a CP violating phase $\delta \simeq \phi ^{\prime }/2$. On the
other hand, the other four possibilities yield $s_{13}\simeq \frac{m_{2}}{%
2m_{3}}\sin 2\theta _{sol}\simeq 0.08$ and a phase $\delta \simeq \phi
^{\prime }$, for the textures with zeros in the first and third columns, or $%
\delta \simeq \phi ^{\prime }+\pi $, for the textures with zeros in the
first and second columns.

\begin{table}
\begin{center}
\begin{tabular}{ccccccccc} \hline
&Texture for ${\bf Y}_{\nu}$ 
&& $|({\bf Y}^{\dagger}_{\nu} \log\frac{M_X}{M} {\bf Y}_{\nu})_{12}|$
&& $|({\bf Y}^{\dagger}_{\nu}\log\frac{M_X}{M}  {\bf Y}_{\nu})_{13}|$ 
&& $|({\bf Y}^{\dagger}_{\nu}\log\frac{M_X}{M}  {\bf Y}_{\nu})_{23}|$ 
\\ \hline \hline
VII$_1$&$ \pmatrix{
 0     & \times  & \times \cr
\times &    0    & \times}$ 
&& 0 
&& $\frac{M_2 m_2}{\langle H^0\rangle^2}\frac{\sin 2\theta_{sol}}{\sqrt{2}}
\log \frac{M_X}{M_2}$ 
&& $\frac{1}{2}\frac{M_1 m_3}{\langle H^0\rangle^2} \log \frac{M_X}{M_1}$
\\  \hline 
VII$_2$&$ \pmatrix{
\times &    0   & \times \cr
   0   & \times & \times}$ 
&& 0
&& $\frac{M_1 m_2}{\langle H^0\rangle^2}\frac{\sin 2\theta_{sol}}{\sqrt{2}}
\log \frac{M_X}{M_1}$
&& $\frac{1}{2}\frac{M_2 m_3}{\langle H^0\rangle^2} \log \frac{M_X}{M_2}$
\\  \hline 
VIII$_1$&$ \pmatrix{
 0     & \times  & \times \cr
\times & \times  &    0    }$
&& $\frac{M_2 m_2}{\langle H^0\rangle^2}\frac{\sin 2\theta_{sol}}{\sqrt{2}}
\log \frac{M_X}{M_2}$
&& 0
&& $\frac{1}{2}\frac{M_1 m_3}{\langle H^0\rangle^2}\log \frac{M_X}{M_1}$
\\  \hline 
VIII$_2$ &$\pmatrix{
\times & \times  &    0   \cr
 0     & \times  & \times}$ 
&& $\frac{M_1 m_2}{\langle H^0\rangle^2}\frac{\sin 2\theta_{sol}}{\sqrt{2}}
\log \frac{M_X}{M_1}$
&& 0
&& $\frac{1}{2}\frac{M_2 m_3}{\langle H^0\rangle^2}\log \frac{M_X}{M_2}$
\\ \hline 
\end{tabular}
\end{center}
\caption{\protect\footnotesize Predictions related 
to rare lepton decay processes}
\label{Table2}
\end{table}

These four textures cannot be discriminated only with neutrino oscillation
experiments. However, if the prediction for the 
CHOOZ angle is confirmed it would be
interesting to determine which is the actual Yukawa matrix at high energies.
As we discussed in the introduction, under some well motivated hypothesis
on the theory at high energies, rare processes and leptogenesis provide
additional information about the neutrino Yukawa matrices.
To quantify this we will assume, as is usually done, that all the soft SUSY
breaking terms are flavour diagonal at high energies, and that the only
particles present in the spectrum are the MSSM particles and the two
right-handed neutrino superfields. Under these assumptions, those four
allowed textures could be discriminated through their predictions for rare
processes. The texture VII yields, at leading order,
a vanishing rate for $\mu \rightarrow
e\gamma $. On the other hand, the matrix with texture VII$_1$ gives 
different predictions for $\tau
\rightarrow e\gamma $ and $\tau \rightarrow \mu \gamma $ than the 
one with texture VII$_2$.  Analogously, neutrino
Yukawa matrices with texture VIII yield vanishing rates
for $\tau \rightarrow e\gamma$, but different predictions
for $\mu \rightarrow e\gamma$ and $\tau \rightarrow \mu \gamma$.
All these results are summarized in Table \ref{Table2}.

The analysis for the remaining allowed texture, texture IV, is more 
complicated, since the matrix
with strict texture zeros would require non perturbative Yukawa couplings
to reproduce the low energy masses and mixing angles. If one keeps track
of these texture zeros in the analysis, one can see that 
the combination $|({\bf Y}_{\nu}^{\dagger}\log 
\frac{M_X}{M} {\bf Y}_{\nu})_{23}|$ diverges
as $M_1/{{\bf Y}^{2}_{\nu}}_{11}$ or $M_2/{{\bf Y}^{2}_{\nu}}_{21}$, 
depending on which of them is larger. However, 
this texture would still predict $U_{13} 
\simeq \pm i \sqrt{\frac{m_2}{m_3}} U_{12}$
as long as $\frac{{{\bf Y}^2_{\nu}}_{11}}{M_1}+
\frac{{{\bf Y}^2_{\nu}}_{21}}{M_2}
\ll \frac{m_2}{\langle H^0 \rangle^2}$, and all 
the Yukawa couplings would remain
perturbative provided the right-handed masses are not too large
($M_2 \lsim \frac{\langle H^0 \rangle^2}{m_3}$). It is interesting to note that
the possibility 
$|({\bf Y}_{\nu}^{\dagger}\log \frac{M_X}{M} 
{\bf Y}_{\nu})_{23}|\simeq 1$, is not
excluded, and would yield rates for $\tau \rightarrow \mu \gamma$ 
at the reach of future experiments, while preserving the prediction for the 
CHOOZ angle. Finally, one can check that 
when $\frac{{{\bf Y}^2_{\nu}}_{11}}{M_1}$ and 
$\frac{{{\bf Y}^2_{\nu}}_{21}}{M_2}$
are sufficiently small to keep the prediction 
$U_{13} \simeq \pm i\sqrt{\frac{m_2}{m_3}} U_{12}$
approximately valid, one obtains
\begin{equation}
|({\bf Y}_{\nu}^{\dagger}\log \frac{M_X}{M} {\bf Y}_{\nu})_{12}|
\simeq |({\bf Y}_{\nu}^{\dagger}\log \frac{M_X}{M} {\bf Y}_{\nu})_{13}|
\simeq \frac{M_k \sqrt{m_2 m_3}}{\langle H^0 \rangle^2} 
\frac{\sin\theta_{sol}}{\sqrt{2}}\log\frac{M_X}{M_k}, 
\end{equation}%
where $k=2$ if ${{\bf Y}^{2}_{\nu}}_{21}\gg {{\bf Y}^{2}_{\nu}}_{11}$, 
and $k=1$ otherwise.

The only unknown parameters in $|({\bf Y}_{\nu}^{\dagger}\log 
\frac{M_X}{M} {\bf Y}_{\nu})_{ij}|$, the 
combination relevant for radiative corrections, are the 
right-handed neutrino masses. Therefore, for each texture, 
one could set constraints on these masses from the bounds on the
rates of rare decays. 
Unfortunately, the constraints are very weak: for typical
values of the soft SUSY masses and $\tan \beta = 3$ one obtains, 
from the present bounds on $\mu \rightarrow e \gamma$, that 
$M_1 \lsim 9 \times 10^{13}$ GeV for the texture VIII$_2$, and 
$M_2 \lsim 9 \times 10^{13}$ GeV for VIII$_1$. These bounds could
be improved by one order of magnitude with the next generation
of experiments, that expect to reach a sensitivity on the
branching ratio of $10^{-13}$. From $\tau \rightarrow \mu \gamma$
or $\tau \rightarrow e \gamma$ the upper bounds on the right-handed
masses are even weaker, of the order of $10^{15}$ GeV. 
Moreover, theoretical
guesses of the right-handed neutrino masses, coming from
scenarios of thermal leptogenesis or particular models, would yield rates
for the rare processes which are too small to be observed in the
next generations of experiments, if radiative corrections induced
by right-handed neutrinos are the only source of flavour violation
in the slepton sector.  

The high predictivity of these textures also allows us to 
obtain information about the CP asymmetry generated in the 
decay of the lightest right-handed neutrino and the parameter 
$\widetilde{m}_1$, which are important for thermal leptogenesis
and some models of non-thermal leptogenesis.
For the texture IV, leptogenesis is not likely to occur since washout 
effects are very large (in this case, $\widetilde{m}_1$ diverges
as ${{\bf Y}^{-2}_{\nu 11}}$ or ${{\bf Y}^{-2}_{\nu 21}}$).
The results for the remaining four allowed textures 
are shown in Table \ref{Table3}.
These expressions for the CP asymmetry are all written in terms
of parameters that are in principle 
measurable at low energies  (the solar angle, 
neutrino masses and the Majorana phase), plus the lightest
right-handed neutrino mass. As we have just
discussed, this mass is related to some matrix elements of
$({\bf Y}^{\dagger}_{\nu} \log \frac{M_X}{M} {\bf Y}_{\nu})$ 
and consequently to the rates for rare decays
in certain scenarios of supersymmetry breaking. 
Therefore, leptogenesis parameters can be written only in
terms of quantities that are, in principle, measurable at low energies.
The explicit expressions are shown in Table \ref{Table3}.
As before, given the expected range of values for the CP asymmetry
in thermal leptogenesis, one could obtain bounds on the rates of 
the rare decays. Again the constraints are very weak: for example,
for the value of the CP asymmetry hinted at by thermal leptogenesis,
$|\epsilon_1| \sim 10^{-6}$, one obtains branching ratios of the order
of $10^{-18}$. If the only source of lepton flavour violation
are the neutrino Yukawa couplings, the observation
of any of these rare processes at rates larger than this would
imply an overproduction of baryon asymmetry in the Universe 
in the two right-handed neutrino scenario.

\begin{table}
\begin{center}
\begin{tabular}{cp{2.5cm}cp{5cm}cc}\hline
&Texture for $Y_{\nu}$ 
&& leptogenesis parameters
&& connection leptogenesis-low energies
\\  \hline 
VII$_1$&$\pmatrix{
 0      & \times    & \times \cr
\times  &    0      & \times}$ 
&& 
\begin{tabular}{c}
$\frac{\epsilon_1}{|\epsilon_{max}|} \simeq  
 \frac{m_2}{m_3} \cos^2\theta_{12} \sin \phi'$ \\
$\widetilde{m}_1 \simeq m_3$
\end{tabular} 
&& $|\epsilon_1| \propto  
\frac{m_2}{m_3} \cos^2\theta_{12} |\sin \phi'| 
\sqrt{BR(\tau \rightarrow \mu \gamma)}$
\\ \hline
VII$_2$&$\pmatrix{
\times &    0    & \times \cr
 0     & \times  & \times}$ 
&& 
\begin{tabular}{c}
$\frac{\epsilon_1}{|\epsilon_{max}|}
\simeq - 
\frac{\cos^2\theta_{12}}{1+\cos^2\theta_{12}} \sin \phi' $ \\
$\widetilde{m}_1 \simeq m_2(1+\cos^2\theta_{12})$
\end{tabular}
&& 
$|\epsilon_1| \propto 
\frac{\cos^2\theta_{12}}{1+\cos^2\theta_{12}}|\sin \phi'|
\sqrt{BR(\tau \rightarrow e \gamma)}$  
\\  \hline 
VIII$_1$&$ \pmatrix{
 0     & \times  & \times \cr
\times & \times  &    0  }$ 
&& 
\begin{tabular}{c}
$\frac{\epsilon_1}{|\epsilon_{max}|} \simeq  
 \frac{m_2}{m_3} \cos^2\theta_{12} \sin \phi'$ \\
$\widetilde{m}_1 \simeq m_3$
\end{tabular} 
&& $|\epsilon_1| \propto  
\frac{m_2}{m_3} \cos^2\theta_{12} |\sin \phi'| 
\sqrt{BR(\tau \rightarrow \mu \gamma)}$
\\  \hline 
VIII$_2$&$\pmatrix{
\times & \times  &    0   \cr
 0     & \times  & \times}$ 
&& 
\begin{tabular}{c}
$\frac{\epsilon_1}{|\epsilon_{max}|} \simeq  -
 \frac{\cos^2\theta_{12}}{1+\cos^2\theta_{12}}\sin \phi'$\\
$\widetilde{m}_1 \simeq m_2(1+\cos^2\theta_{12})$
\end{tabular}
&& $|\epsilon_1| \propto 
\frac{\cos^2\theta_{12}}{1+\cos^2\theta_{12}}|\sin \phi'|
\sqrt{BR(\mu \rightarrow e \gamma)}$
\\ \hline 
\end{tabular}
\end{center}
\caption{\protect\footnotesize Leptogenesis parameters 
written in terms of low energy
observables. The proportionality constant that relates $|\epsilon_1|$ with
the branching ratios for the rare processes depends 
on supersymmetric parameters, and can be read from eq.(\ref{BR}).}
\label{Table3}
\end{table}


\subsection{The general case}

We consider now the effect of $U_l$ and $V_{\nu}$. 
Under the hypotheses on the charged lepton sector explained in 
section 5, it is clear from eqs.(\ref{same-row1}) and (\ref{same-row2})
that in the general case the textures I, II and III
with texture zeros in the same row are still excluded.
Similarly the textures V, VI and IX are also excluded. 
On the other hand texture IV leads to the relation
\bea
U_{13} \simeq -e^{i\alpha_1} \sqrt{\frac{m_e}{m_{\mu}}} U_{23} 
\pm i \sqrt{\frac{m_2}{m_3}} U_{12}
\eea
yielding $0.028\lsim |U_{13}|\lsim 0.13$.

Textures VII yield the prediction for the CHOOZ angle 
\bea
U_{13} \simeq -e^{i\alpha_1} \sqrt{\frac{m_e}{m_{\mu}}} U_{23}
- \frac{m_2}{m_3} \frac{U_{12}U_{22}}{U_{23}}
-e^{i \beta} \sqrt{\frac{m_2}{m_3}} \sqrt{\frac{M_2}{M_1}} 
\sin \omega \cos \omega \frac{U^*_{31}}{U_{23}}, 
\eea
where $\beta$ is an unknown phase, in which definition 
we have also absorbed the indeterminacy coming from the two
possible ``reflections'' in the matrix $R$.
On the other hand, the textures VIII give
\bea
U_{13} \simeq -e^{i\alpha_1} \sqrt{\frac{m_e}{m_{\mu}}} U_{23}
- \frac{m_2}{m_3} \frac{U_{32}U_{12}}{U_{33}}
+ e^{i \beta} \sqrt{\frac{m_2}{m_3}} \sqrt{\frac{M_2}{M_1}} 
\sin \omega \cos \omega \frac{U^*_{21}}{U_{33}}.
\eea

\begin{figure}  
\centerline{
\psfig{figure=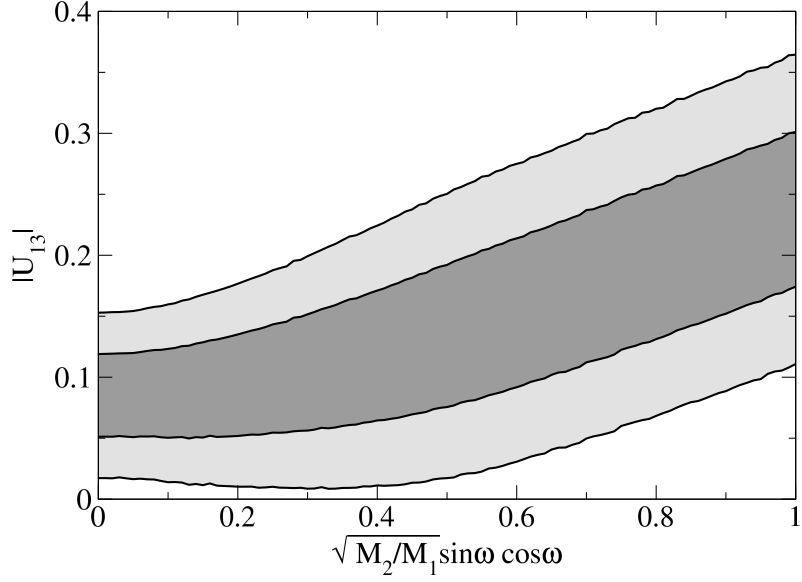,height=12cm,angle=-90}}
\caption{\protect\footnotesize Prediction for the CHOOZ for textures
VII and VIII, allowing contributions to the mixing from the 
right-handed sector (encoded in $\sqrt{M_2/M_1} \sin \omega \cos \omega$)
and the charged lepton sector.} 
\label{fig4}
\end{figure}

The predictions for the CHOOZ angle are very similar for textures VII and VIII,
because of the large angles in the atmospheric and the solar
sector. In Figure \ref{fig4} we show the predictions for the Yukawa matrix
with texture VII$_1$ as 
a function of $\sqrt{M_2/M_1} \sin \omega \cos \omega$. In this plot,
we assign random numbers to the unmeasured phase $\phi'$ and to 
the unknown phase $\beta$, and show the regions at 
1$\sigma$ (darkest) and 2$\sigma$ (lightest) from the main value.
We prefer to leave $\omega$ as
a completely free parameter, since we do not have any hint about the
Majorana sector and we do not know whether the mixing angle $\omega$
is large or small. If the right-handed mass matrix is hierarchical, one
expects this mixing angle to be small, unless some fine-tuning
is taking place. To be precise, one expects $\sin \omega \cos \omega 
\leq \sqrt{M_1/M_2}$, being the bound saturated when there is 
a zero in the (1,1) position of the right-handed mass matrix. So,
in this case the allowed parameter space is 
$0 \leq \sqrt{M_2/M_1} \sin \omega \cos \omega \leq 1$. 
On the other hand, it could happen that the eigenvalues
in the Majorana mass matrix are quasi-degenerate and
the mixing angles could be large without any fine-tuning.
If this is the case, it also happens that $0 \leq \sin \omega \cos \omega 
\lsim \sqrt{M_1/M_2} \leq 1$. The plot shown in Figure \ref{fig4} covers all
the possible natural structures in the right-handed mass matrix.

\begin{figure}  
\centerline{
\psfig{figure=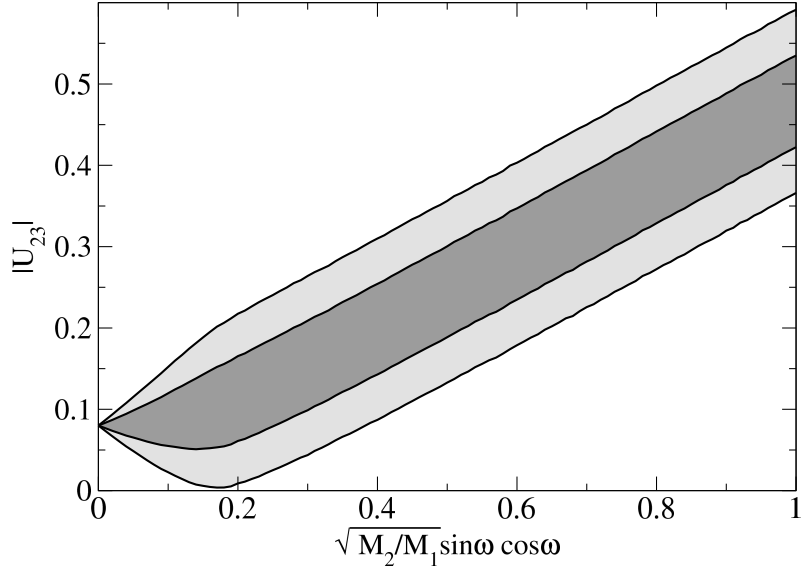,height=12cm,angle=-90}}
\caption{\protect\footnotesize Prediction for the element (2,3) of the MNS
matrix for the textures IX, allowing contributions to the mixing from the 
right-handed sector (encoded in $\sqrt{M_2/M_1} \sin \omega \cos \omega$)
and the charged lepton sector.} 
\label{fig5}
\end{figure}

Finally, for the texture IX  the prediction is
\bea
U_{23}\simeq -\frac{m_2}{m_3}\frac{U_{22}U_{32}}{U_{33}} 
+e^{i \beta} \sqrt{\frac{m_2}{m_3}} \sqrt{\frac{M_2}{M_1}}
\sin\omega\cos\omega \frac{U_{11}}{U_{33}}.
\eea
The numerical results for this case are shown in Figure \ref{fig5}. The inclusion
of the right-handed mixing effects is not enough to reproduce the 
experimental value $|U_{23}|\simeq 1/\sqrt{2}$. Therefore, this 
texture is disfavoured at the 2$\sigma$ level.

The conclusions about the rates of rare decays and leptogenesis
are qualitatively similar to the case in which 
$U_l \simeq 1$ and $V_{\nu} \simeq 1$, namely that the rare processes
are observable and the CP asymmetry is large enough only when
the right-handed masses are large.

\section{Conclusions}\label{conc}
In the case that there are only two right-handed neutrinos or, 
in the three neutrino case, that the heaviest one decouples, 
leads to relations amongst observable properties of neutrinos 
\cite{Ibarra:2003xp}. Here we have presented a general analysis 
of this possibility. In this case the number of parameters involved 
in the  see-saw mechanism is significantly reduced from 6 to 2 
leading to some general phenomenological implications. In particular, 
adequate baryogenesis through thermal leptogenesis is problematic 
due to large washout processes and we give an upper bound on the 
magnitude of the asymmetry. Inhibiting these, requires a large 
right handed neutrino mass which causes problems in supersymmetric
theories due to excessive gravitino production. 
In the 2RHN case there are also strong constraints between the 
level of leptogenesis and lepton flavour violating processes. 
Requiring that these latter processes be acceptable puts an 
upper bound on the heaviest of the right-handed neutrino masses.

Further phenomenological implications apply if the two remaining 
parameters are constrained due to some underlying symmetry of the theory. 
Such symmetries can give rise to anomalously small elements, 
\textquotedblleft texture zeros", in the matrix of Yukawa couplings.  
We have made a complete study of the cases that there are one or two 
such texture zeros in neutrino sector of Yukawa couplings. For the
 case of one texture zero we find constraints on the thermal 
leptogenesis parameters. The case of texture zeros in the 
(1,1), (2,2) and (2,3) positions give small asymmetries, below 
the model independent bound found for the general case. 
The case of texture zeros in the (2,1), (1,2) or (1,3) 
is more promising, nearly saturating the upper bound. 
An interesting issue in these cases is the connection 
between leptogenesis and the low energy phases and we 
have determined this in the cases that the contribution to 
mixing from the charged lepton and Majorana sectors are small.

For the case of two texture zeros we find a prediction for the 
CHOOZ angle and we have classified all possible cases and 
identified the viable ones. The various viable possibilities 
cannot be distinguished on the basis of the CHOOZ angle 
alone but we point out that it may be possible to do so 
from lepton flavour violating processes. For the viable 
cases we have also determined the connection between 
leptogenesis and the low energy phases, again under 
the assumption that the contribution to mixing from the 
charged lepton and Majorana sectors are small. Finally we 
considered the more general case in which the contribution 
to mixing from the charged lepton and Majorana sectors is 
non-negligible. Although less predictive we are still able 
to eliminate several possibilities. For the remainder we 
determined the expected range of the CHOOZ angle consistent 
with the texture zero structure.

\section*{Acknowledgements}
This work was partly funded by the PPARC rolling grant 
PPA/G/O/2002/00479 and the EU network \textquotedblleft 
Physics Across the Present Energy Frontier", HPRV-CT-2000-00148.

\end{document}